\documentclass[aps,prl,preprint,superscriptaddress]{revtex4}

\usepackage{graphicx}
\begin{document}


\title{Heat Flow and Efficiency in a Microscopic Engine}



\author{Baoquan  Ai$^{a}$}\author{Liqiu Wang$^{a}$}
\author{Huizhang Xie$^{b}$}
\author{Lianggang Liu$^{c}$}

\affiliation{$^{a}$ Department of Mechanical Engineering, The University of Hong Kong, Pokfulam Road, Hong Kong.\\
  $^{b}$Department of Physics, South
China University of technology, GuangZhou, China.\\
$^{c}$ Department of Physics, ZhongShan University, GuangZhou,
China,}


\begin{abstract}
\indent We study the energestics of a thermal motor driven by
temperature differences, which consists of Brownian particles
moving in a sawtooth potential with an external load where the
viscous medium is alternately in contact with hot and cold heat
reservoir along space coordinate. The motor can work as a heat
engine or a refrigerator under different conditions. The heat flow
via both potential and kinetic energy are considered. The former
is reversible when the engine works quasistatically and the
latter is always irreversible. The efficiency of the heat engine
(Coefficient Of Performance (COP) of a refrigerator) can never
approach Carnot efficiency (COP).

\end{abstract}

\pacs{05. 40. -a, 05. 70. -a, 87. 10. +e}

\maketitle

\section {Introduction}
\indent Recently, Brownian ratchets (motors) have attracted
considered attention simulated by research on molecular motors
\cite{1,2}. A Brownian ratchet , which appeared in Feynman's
famous textbook for the first time as a thermal ratchet \cite{3},
is a machine which can rectify thermal fluctuation and produce a
directed current. These subjects are motivated by the challenge to
understand molecular motors\cite{4}, nanoscale friction\cite{5},
surface smoothening \cite{6}, coupled Josephson junctions
\cite{7}, optical ratchets and directed motion of laser cooled
atoms \cite{8}, and mass separation and trapping schemes at the
microscale\cite{9}.

\indent The Brownian ratchet are spatially asymmetric but periodic
structure in which transport of Brownian particles is induced by
some nonequilibrium processes \cite{10,11,12,13}, such as external
modulation of the underling potential or a nonequilibrium chemical
reaction coupled to a change of the potential, or contact with
reservoirs at different temperatures. The most intensively studied
quantity has been on the velocity of the transport particle.
However, another important quantity is efficiency of energy
conversion characteristic the operation of the system when the
transported particle does work.

\indent  How efficiency can Brownian ratchets work ? This question
is important not only for the construction of theory of molecular
motors \cite{14} but also for foundation of non-equilibrium
statistical physics. Like Carnot cycle, Brownian heat engine can
extract work from the temperature difference of the heat baths,
where Brownian working material operates as a transducer of
thermal energy into work. The study of the energetics of Brownian
ratchets is relevant for several reasons \cite{15}. Firstly,
highly efficient motors are desirable in order to decrease the
energy consumption or to decrease the heat dissipation. Secondly,
Brownian motors are related to fundamental problems of
thermodynamics and statistical mechanics, such as the Maxwell
demon and the trade-off between entropy and information. Thirdly,
many models proposed in the literature are based on nonequilibrium
fluctuations without specifying their source. On the other hand,
the study of the energetics of such models required a more precise
formulation, since one has to determine the physical nature of the
external agent and verify that the motor is consistent with the
second law of thermodynamics.

\indent
  Recently, Sekimoto has been unambiguously defined the concept of
the heat at mesoscopic scales in terms of Langevin equation
\cite{16}. He refers to the formalism providing this definition as
stochastic energetics. The essential point of this formalism is
that the heat transferred to the system is nothing but the
microscopic work done by both friction force and the random force
in the Langevin equation. Stochastic energetics has been applied
to the study of thermodynamic processes. Derenyi and Astumian
\cite{17} have studied the efficiency of one-dimensional thermally
driven Brownian engines. They identify and compare the three basic
setups characterized by the type of the connection between the
Brownian particle and the two reservoirs: (1) simultaneous
\cite{3}, (2) alternating in time \cite{18}, and (3)position
dependent \cite{19}. Parrondo and Cisneros \cite{15} has reviewed
the literature the energetics of Brownian motors, distinguishing
between forced ratchet, chemical motors-driven out of equilibrium
by difference of chemical potential, and thermal motors-driven by
temperature differences. In this paper we give a detailed study of
the last motors-thermal motors.

\indent Energetics of the thermal motors-driven by temperature
differences are investigated by some previous works \cite{20,21}.
Asfaw \cite{20}et al. have explored the performance of the motors
under various conditions of practical interest such as maximum
power and optimized efficiency. They found that the efficiency can
approach the Carnot efficiency under the quasistatic limit. The
same results are also obtained in Matsuo and Sasa's work \cite{21}
by stochastic energetics theory. The previous works are limited to
case of heat flow via potential. The present work extends the
study to case of heat flow via both potential and kinetics energy.
The motor can work as a heat engine or a refrigerator under
various conditions. The efficiency of heat engine (COP of
refrigerator) is different from the results of previous works and
can never approach the Carnot efficiency (COP). The heat flow via
potential is reversible when the engine works at quasistatic
limit. The heat flow via the kinetic is always irreversible in
nature.

\section{Thermal motor works as a heat engine}
The model (shown in Fig. 1) consists of Brownian particles moving
in a sawtooth potential with an external load where the medium is
alternately in contact with hot and cold heat reservoirs along the
space coordinate.  Let $E$ be the barrier height of the potential.
$L_{1}$, $L_{2}$ are the length of the left part and the right
part of the ratchet, respectively. $E+FL_{1}$ is the energy needed
for a forward jump, while $E-FL_{2}$ is the energy required for a
backward jump. The left part of a period ratchet is at temperature
$T_{H}$ (hot reservoir)and the right one is at temperature $T_{C}$
(cold reservoir). We can assumes that the rates of both forward
and backward jumps are proportional to the corresponding
Arrhenius' factor \cite{a1}, such that

\begin{figure}[htbp]
\includegraphics[width=8cm,height=6cm]{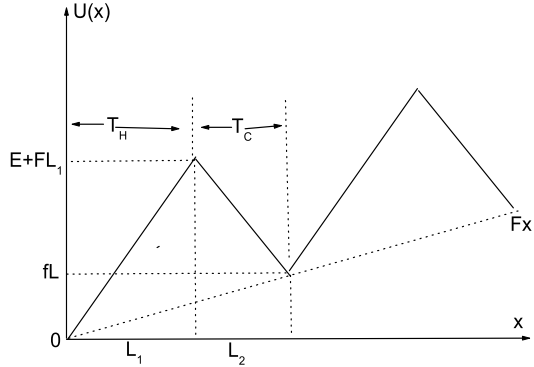}
\caption{Schematic illustration of the potential,
$U(x)=U_{0}(x)+Fx$, $U_{0}(x)$ is a piecewise linear and periodic
potential. $Fx$ is a potential due to the load. The period of the
potential is $L=L_{1}+L_{2}$. The temperature profiles are also
shown.}
\end{figure}

\begin{equation}\label{}
    \dot{N}_{+}=\frac{1}{t}\exp{[-\frac{E+FL_{1}}{k_{B}T_{H}}]},
\end{equation}

\begin{equation}\label{}
    \dot{N}_{-}=\frac{1}{t}\exp{[-\frac{E-FL_{2}}{k_{B}T_{C}}]},
\end{equation}
are the number of forward and backward jumps per unit time,
respectively, $k_{B}$ the Boltzmann constant, $t$ a
proportionality constant with dimensions of time.

 \indent If $\dot{N}_{+}>\dot{N}_{-}$, the ratchet works as a
 two-reservoir heat engine shown in Fig. 2.
\begin{figure}[htbp]
\includegraphics[width=8cm,height=4cm]{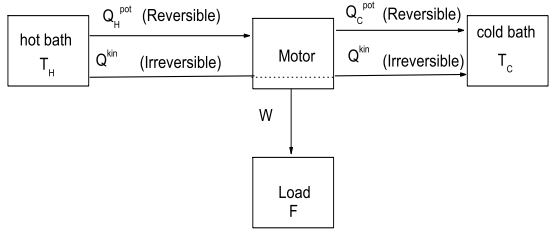}
\caption{The engine acts as a heat engine. Heat flows via both
potential energy and kinetic energy in a thermal motor in contact
with two thermal baths at temperatures $T_{H}>T_{C}$, $W$ is power
output, heat flows via potential energy is reversible, heat flows
via kinetic energy is irreversible.}
\end{figure}
 The rate of heat flow from hot
 reservoir to the heat engine via potential is given by

\begin{equation}\label{}
\dot{Q}_{H}^{pot}=(\dot{N}_{+}-\dot{N}_{-})(E+FL_{1}).
\end{equation}
\indent The rate of heat flow from the engine to the cold
reservoir via potential is
\begin{equation}\label{}
\dot{Q}_{C}^{pot}=(\dot{N}_{+}-\dot{N}_{-})(E-FL_{2}).
\end{equation}

The heat flow via the kinetic energy of the particle is much more
complicated to determined \cite{15}. Whenever a particle stay at a
hot segment (temperature $T_{H}$) it absorbs
$\frac{1}{2}k_{B}T_{H}$ energy on average from the hot reservoir.
It can pick up $\frac{1}{2}k_{B}T_{C}$ energy on average from the
cold reservoir when the particle stay at a cold segment. It is
obvious that when a particle leaves from a hot segment to a cold
segment and then returns to the hot segment, the hot reservoir
will lost $\frac{1}{2}(k_{B}T_{H}-k_{B}T_{C})$ energy on average,
the lost energy is released to the colder reservoir and never to
the hot reservoir or to the particle's potential energy, which
indicates the inherently irreversible nature of this heat flow. On
the other hand, if a particle goes out from a hot segment to a
cold segment and never returns to the hot segment, the hot
reservoir will lost $\frac{1}{2}k_{B}T_{H}$ energy on average.
Therefore the rate of net heat flow via kinetic energy from the
hot reservoir to the cold reservoir is given by
\begin{equation}\label{}
\dot{Q}^{kin}=\frac{1}{2}\dot{N}_{+}k_{B}T_{H}-\frac{1}{2}\dot{N}_{-}k_{B}T_{C}.
\end{equation}
\indent Therefore, the rate of total heat transferred from the hot
reservoir is given by
\begin{eqnarray}\nonumber
  \dot{Q}_{H}&=&\dot{Q}_{H}^{plot}+\dot{Q}^{kin}, \\
  &=&(\dot{N}_{+}-\dot{N}_{-})(E+FL_{1})+\frac{1}{2}\dot{N}_{+}k_{B}T_{H}-\frac{1}{2}\dot{N}_{-}k_{B}T_{C}.
\end{eqnarray}

\indent The rate of total heat transferred to the cold reservoir
is

\begin{eqnarray}\nonumber
  \dot{Q}_{C}&=&\dot{Q}_{C}^{plot}+\dot{Q}^{kin} \\
  &=&(\dot{N}_{+}-\dot{N}_{-})(E-FL_{2})+\frac{1}{2}\dot{N}_{+}k_{B}T_{H}-\frac{1}{2}\dot{N}_{-}k_{B}T_{C}.
\end{eqnarray}

\indent The power output is
\begin{equation}\label{}
    \dot{W}= \dot{Q}_{H}- \dot{Q}_{C}=(\dot{N}_{+}-\dot{N}_{-})FL.
\end{equation}

 It is easy to obtain the efficiency of the heat engine
\begin{equation}\label{}
    \eta=\frac{\dot{W}}{\dot{Q}_{H}}.
\end{equation}
\indent If the heat flow via the kinetic energy is not considered,
the efficiency is given by
\begin{equation}\label{}
    \eta^{pot}=\frac{\dot{W}}{\dot{Q}_{H}^{pot}}.
\end{equation}

In order to discuss simply, we can rewritten Eq. (5-10) in a
dimensionless form, then we get
\begin{equation}\label{}
    q_{H}=\frac{\dot{Q}_{H}t}{k_{B}T_{H}}=e^{-\epsilon/\tau}[(\epsilon+\frac{1}{2}+\mu f)e^{f_{0}-\mu
    f}-(\epsilon+\frac{1}{2}\tau+\mu f)e^{(1-\mu)f/\tau}],
\end{equation}
\begin{equation}\label{}
    q_{C}=e^{-\epsilon/\tau}[(\epsilon+\frac{1}{2}+\mu f-f)e^{f_{0}-\mu
    f}-(\epsilon+\frac{1}{2}\tau+\mu f-f)e^{(1-\mu)f/\tau}],
\end{equation}

\begin{equation}\label{}
    q^{kin}=\frac{\dot{Q}_{kin}t}{k_{B}T_{H}}=\frac{1}{2}e^{-\epsilon/\tau}[e^{f_{0}-\mu
    f}-\tau e^{(1-\mu)f/\tau}],
\end{equation}

\begin{equation}\label{}
    w=\frac{\dot{W}t}{k_{B}T_{H}}=e^{-\epsilon/\tau}[e^{f_{0}-\mu
    f}-e^{(1-\mu)f/\tau}]f,
\end{equation}
\begin{equation}\label{}
    \eta=\frac{[e^{f_{0}-\mu
    f}-e^{(1-\mu)f/\tau}]f}{(\epsilon+\mu f+\frac{1}{2})e^{f_{0}-\mu
    f}-(\epsilon+\mu f+\frac{1}{2}\tau)e^{(1-\mu)f/\tau}},
\end{equation}

\begin{equation}\label{}
    \eta^{pot}=\frac{f}{\epsilon+\mu f},
\end{equation}
where
\begin{equation}\label{}
    f=\frac{FL}{k_{B}T_{H}},\epsilon=\frac{E}{k_{B}T_{H}},\tau=\frac{T_{C}}{T_{H}},
    \mu=\frac{L_{1}}{L}, f_{0}=\frac{(1-\tau)\epsilon}{\tau}.
\end{equation}
\indent From the above equations, one has $\epsilon \geq 0$,
$0\leq \tau \leq 1$, $0 \leq \mu \leq1$. Since the motor acts as a
heat engine, the power output can not be negative, namely, $w \geq
0$. one can also obtain the value range of $f$, $0\leq f\leq
f_{m}$, where
\begin{equation}\label{}
    f_{m}=\frac{(1-\tau)\epsilon}{(\tau-1)\mu+1}.
\end{equation}
\indent Therefore, it is easy to obtain
\begin{equation}\label{}
    \eta^{pot}=\frac{f}{\epsilon+\mu f}\leq \frac{f_{m}}{\epsilon+\mu
    f_{m}}=1-\tau=1-\frac{T_{C}}{T_{H}}=\eta_{C},
\end{equation}
where $\eta_{C}$ is Carnot efficiency. However, $\eta^{pot}$
attains the Carnot efficiency, namely, $f=f_{m}$ which indicates
that the power output is zero. From Eq. (15) and Eq. (16), it
obvious that $\eta$ is always less than $\eta^{pot}$.  The results
are given by Fig. 3-10.


\begin{figure}[htbp]
\includegraphics[width=8cm,height=6cm]{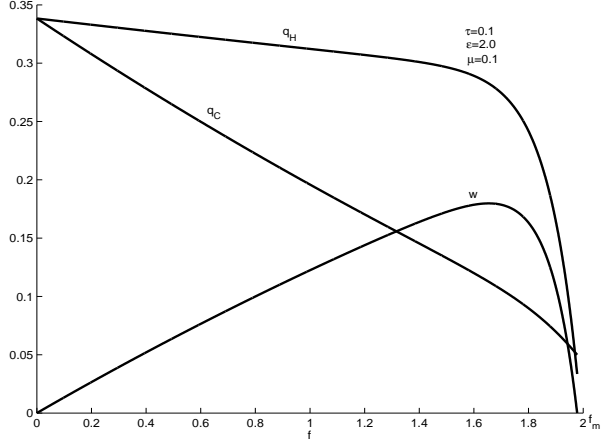}
\caption{Dimensionless heat flow $q_{H}$,$q_{C}$ and power output
$w$ vs the load $f$ at $\tau=0.1$,$\epsilon=2.0$,$\mu=0.1$. }
\end{figure}

\indent Figure 3 shows that the heat flow $q_{H}$,  $q_{C}$ and
power output $w$ as a function of the load $f$. When $f=0$,
namely, the engine runs without a load, $q_{H}$ is equal to
$q_{C}$, which indicates that the heat that absorbs from the hot
reservoir releases to the cold reservoir entirely and no power
output is obtained. When $f$ increases, $q_{H}$ and $q_{C}$
decreases. when $f\rightarrow 0$, no power output is obtained
($w$=0). When $f\rightarrow f_{m}$, no currents occur, the ratchet
can't give any power output. So the power output $w$ has a maximum
value at certain value of $f$ which depends on $\tau$, $\epsilon$
and $\mu$.

\begin{figure}[htbp]
\includegraphics[width=8cm,height=6cm]{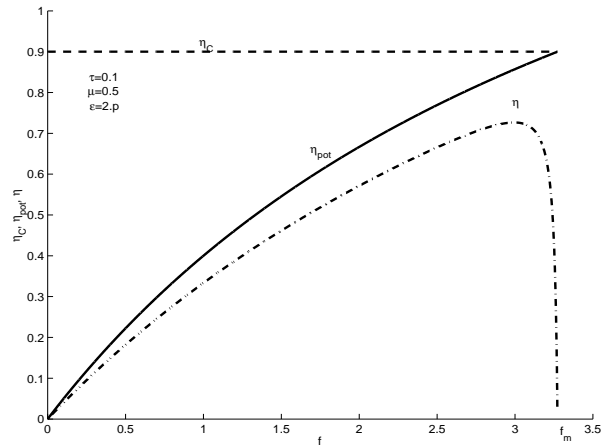}
\caption{$\eta_{C}$, $\eta_{pot}$ and $\eta$ vs the load $f$ at
$\tau=0.1$,$\epsilon=2.0$,$\mu=0.1$.}
\end{figure}

\indent Figure 4 shows the variation of the efficiency $\eta_{C}$,
$\eta_{pot}$, $\eta$ with the load $f$. If the heat flow via
kinetic energy is ignored, the efficiency $\eta_{pot}$ increases
with the load $f$, it approaches the Carnot efficiency $\eta_{C}$
at quasistatic condition ($f$=$f_{m}$). The result is also
presented in Asfaw's work \cite{20}. When the heat flow via
kinetic energy are considered, the curve of the efficiency $\eta$
is observed to be bell-shaped, a feature of resonance. The
efficiency $\eta$ is always less than $\eta_{pot}$ and never
approaches Carnot efficiency $\eta_{C}$. It is obvious that the
heat flow via kinetic energy is always irreversible and energy
leakage is inevitable, so the efficiency is less than $\eta_{pot}$
and can't approach Carnot efficiency.

\begin{figure}[htbp]
\includegraphics[width=8cm,height=6cm]{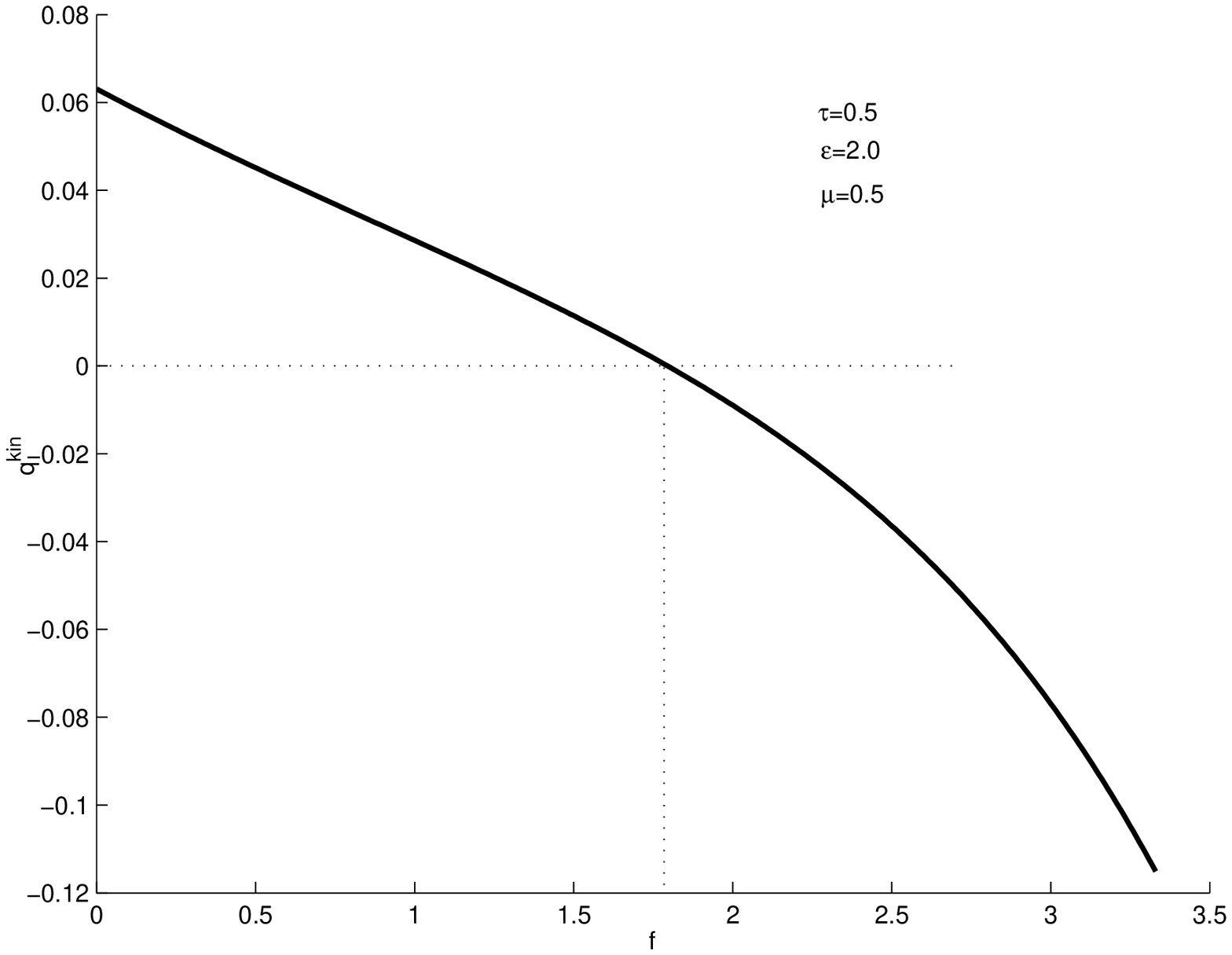}
\includegraphics[width=8cm,height=6cm]{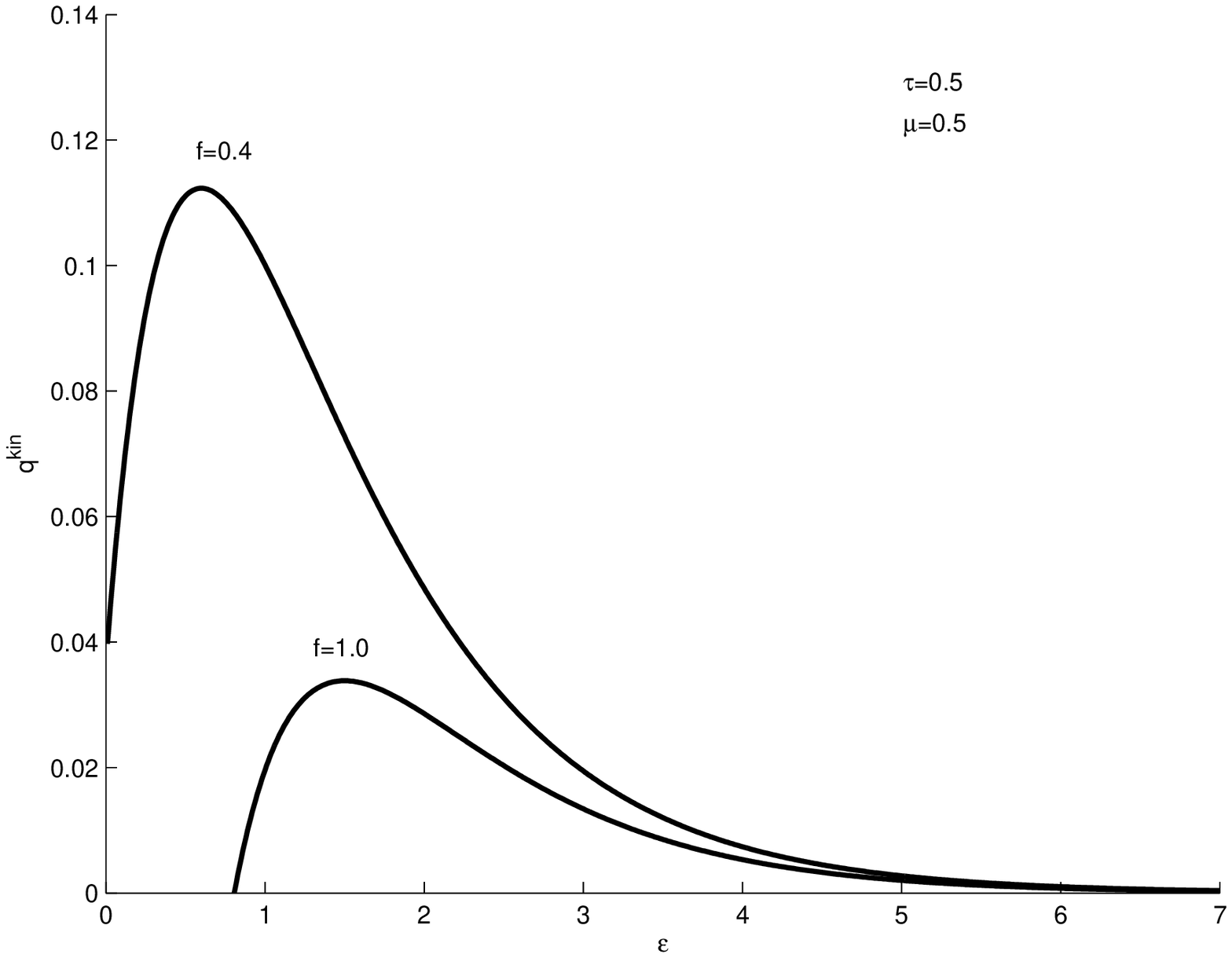}
\caption{(a)Dimensionless heat flow $q^{kin}$ vs the load $f$
($\tau=0.5$, $\epsilon=2.0$, $\mu=0.5$); (b)Dimensionless heat
flow $q^{kin}$ vs barrier height $\epsilon$ for different values
of $f=0.4, 1.0$ at $\tau=0.5$, $\mu=0.5$.}
\end{figure}

\indent Figure 5a shows the heat flow $q^{kin}$ out of hot
reservoir via kinetic energy as a function of the load $f$. When
$f<f_{m}$, $q^{kin}$ is positive. When $f>f_{m}$, $q^{kin}$ is
negative. No heat flow occurs at $f$=$f_{m}$. It is found that the
heat flow via kinetic is dependant on the current of the ratchet.
Figure 5b shows the heat flow out of the hot reservoir via kinetic
energy as a function of barrier height $\epsilon$. The curve is
bell-shaped. Therefore, there is an optimized value of $\epsilon$
at which $q^{kin}$ takes its maximum value.

\begin{figure}[htbp]
\includegraphics[width=8cm,height=6cm]{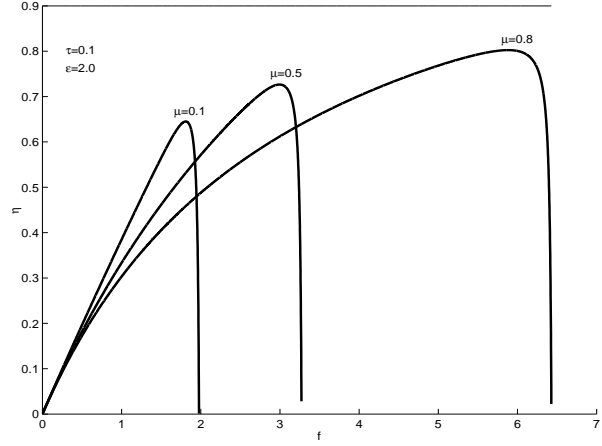}
\caption{Dimensionless power output $w$ vs the load $f$ for
different values of $\mu=0.1, 0.5, 0.8$ at $\tau=0.1$,
$\epsilon=1.0$.}
\end{figure}
\indent Figure 6 shows the power output $w$ as a function of the
load $f$ for different values of $\mu=0.1, 0.5, 0.8$. From the
figure, we can see that the power output has a maximum value at
certain value of $f$.  The maximum load $f_{m}$ of the engine
changes with the parameter $\mu$ of asymmetry in potential. In
other word, the structure of the potential determines the maximum
load capability of the engine.

\begin{figure}[htbp]
\includegraphics[width=8cm,height=6cm]{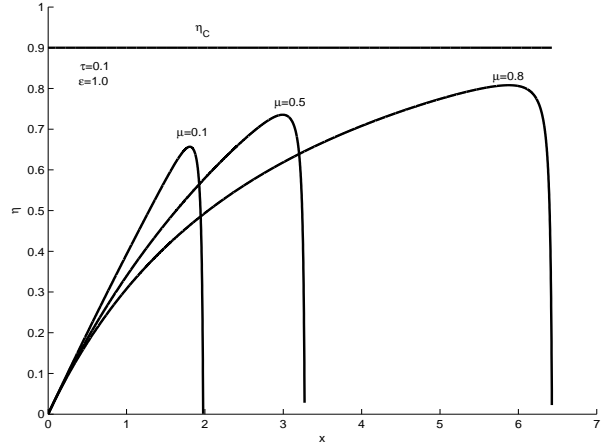}
\caption{Efficiency $\eta$ vs the load $f$ for different values of
$\mu=0.1, 0.5, 0.8$ at $\tau=0.1$, $\epsilon=1.0$.}
\end{figure}
\indent Figure 7 shows the variation of the efficiency $\eta$ with
the load $f$ for different values of $\mu=0.0,0.5,0.8$. The
maximum value of $\eta$ increases with $\mu$. However, the maximum
value of $\eta$ can't approach the Carnot efficiency.

\begin{figure}[htbp]
\includegraphics[width=8cm,height=6cm]{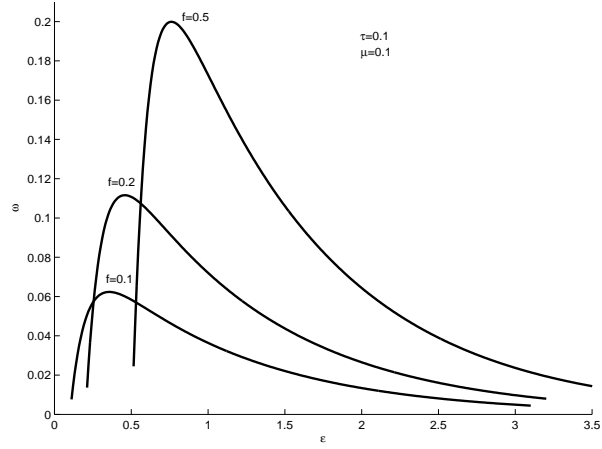}
\caption{Dimensionless power output $w$ vs barrier height
$\epsilon$ for different values of $f$=0.1, 0.2, 0.5 at
$\tau=0.1$, $\mu=0.1$.}
\end{figure}

\indent Figure 8 shows the power output $w$ as a function of the
barrier height $\epsilon$ for different values of the load $f=0.1,
0.2,0.5$. When $\epsilon\rightarrow 0$, the effect of ratchet
disappears, the power output tends to zero. When $\epsilon
\rightarrow +\infty$ so that the particle can't pass the barrier,
the power output $w$ goes to zero, too. The power output $w$ has a
maximum value at certain value of $\epsilon$ which is dependant on
$\tau$, $\mu$ and $f$. On the other hand, the minimum height of
the barrier for working as a heat engine increases with the load
$f$.

\begin{figure}[htbp]
\includegraphics[width=8cm,height=6cm]{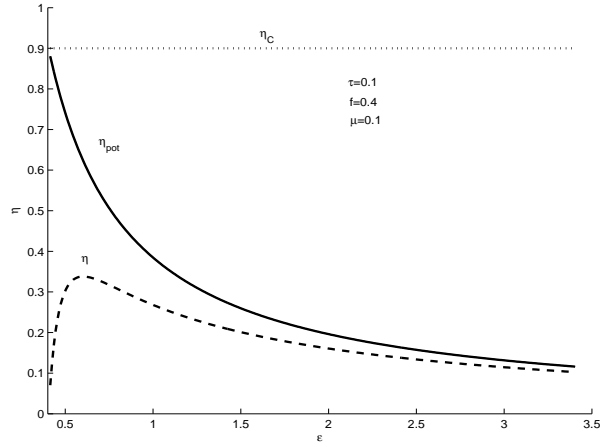}
\caption{$\eta_{C}$, $\eta_{pot}$ and $\eta$ vs the barrier height
$\epsilon$ at $\tau=0.1$, $f=0.4$, $\mu=0.1$.}
\end{figure}

\indent Figure 9 shows the efficiency $\eta_{C}$, $\eta_{pot}$ and
$\eta$ as a function of the barrier height. The efficiency
$\eta_{pot}$ without the heat flow via kinetic energy approaches
the Carnot efficiency at $\epsilon$=$\epsilon_{m}$, at which no
power output occurs. The efficiency $\eta$ is a peaked function of
the barrier height $\epsilon$ which is dependant on values of
$\tau$, $f$ and $\mu$.

\begin{figure}[htbp]
\includegraphics[width=8cm,height=6cm]{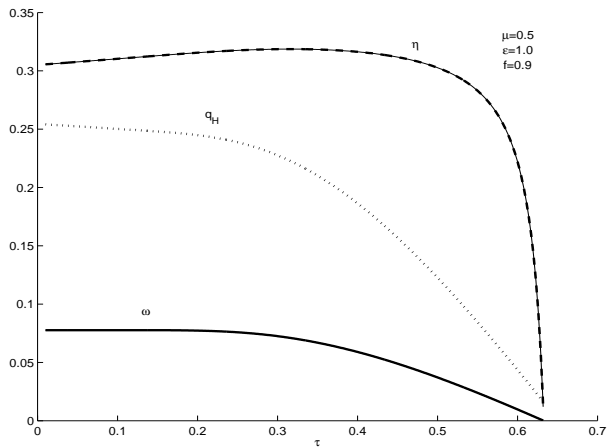}
\caption{Dimensionless heat flow $q_{H}$, $q_{C}$ and power output
$w$ vs $\tau$ at $f=0.9$, $\epsilon=1.0$, $\mu=0.5$.}
\end{figure}

\indent Figure 10 shows plot of $w$, $q_{H}$ and $\eta$ versus
$\tau$. From the figure, we can see that $w$, $q_{H}$ and $\eta$
change very slowly at small $\tau$ and they decreases drastically
near $\tau=\tau_{m}$.

\indent When the load $f$ is less than the maximum load $f_{m}$,
the motor works as a heat engine. The power output is a peaked
function of the load $f$ and the barrier height $\epsilon$. The
efficiency $\eta$ is less than the efficiency $\eta_{pot}$ and can
never approach the Carnot efficiency $\eta_{C}$. The heat flow via
kinetic energy causes the energy leakage unavoidably and reduces
the efficiency of the heat engine.

\section{Thermal motor works as a refrigerator}
\indent If the load is large enough along with appropriately
chosen other quantities, the motor will run in the reverse
direction $\dot{N}_{+} \leq \dot{N}_{-}$. The motor will absorb
the heat from the cold reservoir and release it to the hot
reservoir. Under this condition, the thermal motor acts as a
refrigerator shown in Fig. 11.

\begin{figure}[htbp]
\includegraphics[width=8cm,height=4cm]{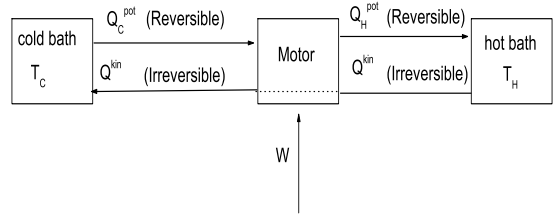}
\caption{The engine acts as a refrigerator. Heat flows via both
potential energy and kinetic energy in a thermal motor in contact
with two thermal baths at temperatures $T_{H}>T_{C}$, $W$ is power
input, heat flow via potential energy is reversible, heat flow
via kinetic energy is irreversible.}
\end{figure}
\indent Similarly, the rate of total heat transferred to the hot
reservoir is given by
\begin{equation}\label{}
    \dot{Q}_{H}=\dot{Q}_{H}^{pot}+\dot{Q}^{kin}=(\dot{N}_{-}-\dot{N}_{+})(E+FL_{1})-\frac{1}{2}k_{B}(\dot{N}_{+}T_{H}-\dot{N}_{-}T_{C}).
\end{equation}
\indent The rate of total heat transferred from the cold reservoir
to the heat engine is given by
\begin{equation}\label{}
    \dot{Q}_{C}=\dot{Q}_{C}^{pot}+\dot{Q}^{kin}=(\dot{N}_{-}-\dot{N}_{+})(E-FL_{2})-\frac{1}{2}k_{B}(\dot{N}_{+}T_{H}-\dot{N}_{-}T_{C}).
\end{equation}
The power input is
\begin{equation}\label{}
    \dot{W}=(\dot{N}_{-}-\dot{N}_{+})FL.
\end{equation}

\indent The above equations can also be rewritten in a
dimensionless form with Eq. (17)
\begin{equation}\label{}
q_{H}=\frac{\dot{Q}_{H}t}{k_{B}T_{H}}=e^{-\epsilon/\tau}[(\epsilon+\frac{1}{2}\tau+\mu
f)e^{(1-\mu)f/\tau}-(\epsilon+\frac{1}{2}+\mu f)e^{f_{0}-\mu f}],
\end{equation}

\begin{equation}\label{}
    q_{C}=\frac{\dot{Q}_{C}t}{k_{B}T_{H}}=e^{-\epsilon/\tau}[(\epsilon+\frac{1}{2}\tau+\mu f-f)e^{(1-\mu)f/\tau}-(\epsilon+\frac{1}{2}+\mu f-f)e^{f_{0}-\mu
    f}],
\end{equation}
\begin{equation}\label{}
    w=\frac{\dot{W}t}{k_{B}T_{H}}=e^{-\epsilon/\tau}[e^{(1-\mu)f/\tau}-e^{f_{0}-\mu
    f}]f.
\end{equation}
\indent For a refrigerator, $\dot{N}_{+}\leq \dot{N}_{-}$, one can
get $f\geq f_{m}$. The heat flow from cold reservoir must not be
negative ($q_{C}\geq 0$), which is important feature of a
refrigerator.

As for a refrigerator, its generalized COP is most important,
which is given by
\begin{equation}\label{}
    P=\frac{q_{C}}{w}=\frac{(\epsilon+\frac{1}{2}\tau+\mu x-x)e^{(1-\mu)x/\tau}-(\epsilon+\frac{1}{2}+\mu x-x)e^{x_{0}-\mu
    x}}{[e^{(1-\mu)x/\tau}-e^{x_{0}-\mu
    x}]x}
\end{equation}
\indent If the heat flow via the kinetic energy is not considered,
the COP is given by
\begin{equation}\label{}
    P_{pot}=\frac{\epsilon}{f}+\mu-1\leq
    \frac{\epsilon}{f_{m}}+\mu-1=\frac{\tau}{1-\tau}=P_{C}
\end{equation}
where $P_{C}$ is COP of a Carnot refrigerator. When $P_{pot}$
attains the $P_{C}$, namely, $f=f_{m}$, the power input is zero.
The results are shown in Fig. 12-14.

\begin{figure}[htbp]
\includegraphics[width=8cm,height=6cm]{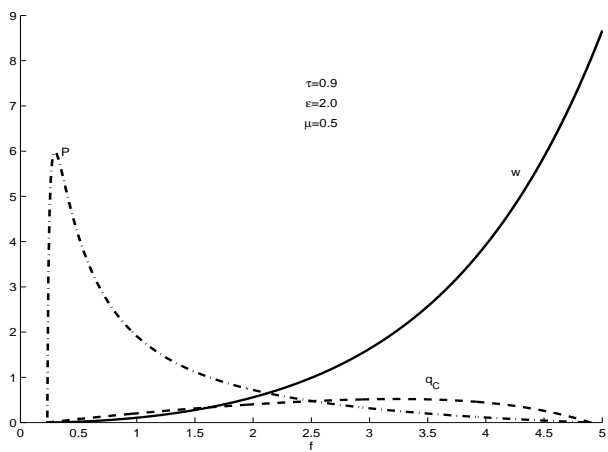}
\caption{Dimensionless heat flow $P$, $q_{C}$ and power input $w$
vs the load $f$ at $\tau=0.9$, $\epsilon=2.0$, $\mu=0.5$.}
\end{figure}
\indent Figure 12 shows $P$, $q_{C}$ and $w$ as a function of the
input force $f$. The power input $w$ increases with the input
force $f$. The heat $q_{C}$ that absorbs from the cold reservoir
has a maximum value at certain value of input force $f$. The COP
$P$ of the refrigerator takes its maximum value at which the power
input tends to zero.

\begin{figure}[htbp]
\includegraphics[width=8cm,height=6cm]{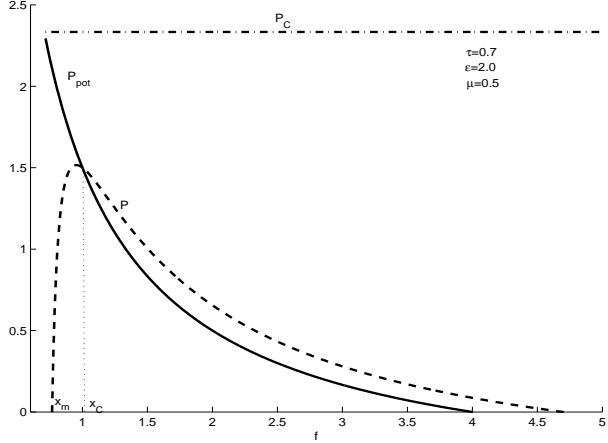}
\caption{$P_{C}$, $P_{pot}$ and $P$ vs input force $f$ at
$\tau=0.7$, $\epsilon=2.0$, $\mu=0.5$.}
\end{figure}
 \indent Figure 13 gives the
plot of COP $P_{C}$, $P_{pot}$ and $P$ versus the input force $f$.
When the heat flow via kinetic energy is ignored, the COP of the
refrigerator $P_{pot}$ decreases with increasing of $f$ and it
approaches the Carnot COP at $f=f_{m}$. When the heat flow via
kinetic energy is considered, the curve of COP $P$ is bell-shaped
and it can never approach the Carnot COP $P_{C}$.  The COP $P$
without the heat flow via kinetic energy is less than $P_{pot}$
when $f_{m}<f<f_{C}$, larger than $P_{pot}$ when $f>f_{C}$ and
equal to $P_{pot}$ at $f=f_{C}$. However, $P$ can never approach
the Carnot COP $P_{C}$.
\begin{figure}[htbp]
\includegraphics[width=8cm,height=6cm]{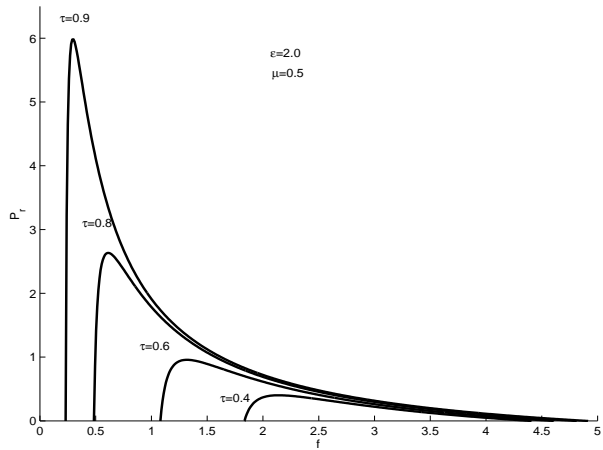}
\caption{The COP $P$ vs the input force $f$ for different values
of $\tau=0.4, 0.6, 0.8, 0.9$ at $\epsilon=0.1$, $\mu=0.5$.}
\end{figure}

\indent Figure 14 shows the variation of the COP $P$ with the
input force $f$ for different values of the $\tau=0.4, 0.6,0.8,
0.9$. From the figure, it is easy to see that the COP $P$
increases with the value of $\tau$ and the minimum input force
$f_{m}$ increases with the decreasing of the value of $\tau$. In
an other word, when temperature difference is small, for example,
$\tau=0.9$, the heat leakage is small, so the COP $P$ is large. On
the contrary, when the minimum input force is large, the
temperature difference is large, for example, $\tau=0.4$.

\indent When $f$ is so large that the motor runs reversibly, $f$
becomes the input force, so the motor works as a refrigerator. The
COP $P$ is a peaked function of input force $f$ which indicates
the the COP can not obtain a maximum value at quasi-static limit.
The COP of refrigerator can never approach the Carnot COP $P_{C}$
because of the heat flow via kinetic energy.
\section{Concluding Remarks}
\indent In present work, we study the energestic of a thermal
motor which consists of Brownian particles moving a sawtooth
potential with an external load where the viscous medium is
alternately in contact with hot and cold reservoirs along the
space coordinate. The thermal motor can works as both a heat
engine and a refrigerator. We make a clear distinction between the
heat flow via the kinetic and the potential energy of the
particle, and show that the former is always irreversible and the
latter may be reversible when the engine works quasistatically.

\indent When the external load is not enough, the thermal motor
can work as a heat engine. The power output has a maximum value at
certain value of the load $f$ which dependant on the others
parameters. If only the heat flow via potential is considered, the
efficiency $\eta_{pot}$ increases with the load $f$ and approaches
the Carnot efficiency $\eta_{C}$ under quasistatic condition,
which agrees with the previous work. When the heat flow via
kinetic energy is also considered, the efficiency $\eta$ is less
than $\eta_{pot}$ and can never approach the Carnot efficiency
$\eta_{C}$. It is also found that the structure of the potential
decides the maximum load capability of the heat engine. The heat
flow via potential is reversible, while the heat flow via kinetic
energy is irreversible. The heat flow via kinetic energy reduces
the heat engine efficiency.

\indent When the external load is so large that the motor moves
reversely, the thermal motor can work as a refrigerator. When the
heat flow via kinetic energy is ignored, the COP $P_{pot}$
decreases with increasing of the input force $f$ and can approach
the Carnot COP $P_{C}$. When the heat flow via both potential and
kinetic energy are considered, the COP $P$ is less than $P_{pot}$
when $f_{m}<f<f_{C}$, larger than $P_{pot}$ when $f>f_{C}$ and
equal to $P_{pot}$ at $f=f_{C}$. However, $P$ can never approach
the Carnot COP $P_{C}$.

\indent
 The thermal motor was initially proposed in an attempt of
 describing molecular motors in biological systems. The heat flow via
 energy is irreversible, so the efficiency (COP) can not approach the
 the Carnot efficiency (COP) at a quasistatic condition. However, Molecular
 motors are known to operator efficiency \cite{d1,d2,d3,d4,d5}. They
 can convert molecular scale chemical energy into macroscopic
 mechanical work with high efficiency in water at room
 temperatures, where the effect of thermal fluctuation is
 unavoidable.  Thus, the next challenging question would be " How to
reduce the heat flow via kinetic energy ? "

{\bf Acknowledgements}\\
 \indent The financial support from the outstanding Young Researcher Award 2000-2001 and the CRCG, the University of Hong Kong to L Wang is gratefully acknowledged.\\

\end{document}